\documentclass[12pt, a4paper]{article}
\usepackage[breaklinks,citecolor=black,urlcolor=black]{hyperref}
\usepackage{booktabs} 
\usepackage[font=small,labelfont=bf]{caption} 
\usepackage{amsfonts, amsmath, amsthm, amssymb}
\usepackage{indentfirst} 
\usepackage{algorithmic}
\usepackage{ascmac}
\usepackage{url}
\usepackage{breakurl}
\usepackage{threeparttable}

\usepackage{graphicx}
\begin{document}

\title{A positive-definiteness-assured block Gibbs sampler for Bayesian graphical models with shrinkage priors\thanks{This research is supported by JSPS Grants-in-Aid for Scientific Research Grant Number 19K01592 and the Keio Economic Society.}}
\author{Sakae Oya\footnote{\url{sakae.prosperity21@keio.jp}
}\\Graduate School of Economics, Keio University\\and\\ Teruo Nakatsuma\\Faculty of Economics, Keio University}
\date{October, 2020}
\maketitle

\begin{abstract}
Although the block Gibbs sampler for the Bayesian graphical LASSO proposed by Wang (2012) has been widely applied and extended to various shrinkage priors in recent years, it has a less noticeable but possibly severe disadvantage that the positive definiteness of a precision matrix in the Gaussian graphical model is not guaranteed in each cycle of the Gibbs sampler. Specifically, if the dimension of the precision matrix exceeds the sample size, the positive definiteness of the precision matrix will be barely satisfied and the Gibbs sampler will almost surely fail. In this paper, we propose modifying the original block Gibbs sampler so that the precision matrix never fails to be positive definite by sampling it exactly from the domain of the positive definiteness. As we have shown in the Monte Carlo experiments, this modification not only stabilizes the sampling procedure but also significantly improves the performance of the parameter estimation and graphical structure learning. We also apply our proposed algorithm to a graphical model of the monthly return data in which the number of stocks exceeds the sample period, demonstrating its stability and scalability.

\end{abstract}

\section{Introduction}
Suppose $\boldsymbol{Y}$ is an $(n\times p)$ data matrix of $p$ variables and $n$ observations and the $t$-th row vector of $\boldsymbol{Y}$, $\boldsymbol{y}_{t}$ $(1\leqq t\leqq n)$, follows a multivariate normal distribution $\mathcal{N}(\boldsymbol{0},\boldsymbol{\Omega}^{-1})$, where $\boldsymbol{\Omega}=(\omega_{ij})$, $(1\leqq i, j\leqq p)$ is the inverse of the covariance matrix, called the precision matrix. In the multivariate normal distribution, $\omega_{ij}=0$ implies that $y_{ti}$ and $y_{tj}$ are independent. Therefore, a set of nonzero off-diagonal elements in $\boldsymbol{\Omega}$ constitutes an undirected graphical structure among $(y_{t1},\dots,y_{tp})$ that is called the Gaussian graphical model.

We may estimate $\boldsymbol{\Omega}$ by maximizing the log likelihood:
\begin{equation}
\label{ggm.logl}
 \ell(\boldsymbol{\Omega}) = -\frac{np}2\log 2\pi
 +\frac{n}2\log|\boldsymbol{\Omega}|-\frac12\mathrm{tr}\left(\boldsymbol{S}\boldsymbol{\Omega}\right),
\end{equation}
where $\boldsymbol{S}=(s_{ij})=\boldsymbol{Y}^{\intercal}\boldsymbol{Y}$. In practice, however, a maximum likelihood estimator (MLE) with \eqref{ggm.logl} does not produce estimates of off-diagonal $\omega_{ij}$'s that are exactly equal to zero. To obtain ``zero estimates'' of $\omega_{ij}$'s, we may employ a LASSO-type penalized MLE:
\begin{equation}
\label{glasso}
\max_{\boldsymbol{\Omega}\in M^+} \frac{n}2\log|\boldsymbol{\Omega}| - \frac12\mathrm{tr}\left(\boldsymbol{S}\boldsymbol{\Omega}\right) - \lambda\Vert\boldsymbol{\Omega}\Vert_1,
\end{equation}
where $\Vert\boldsymbol{\Omega}\Vert_1 = \sum_{i\leqq j}|\omega_{ij}|$ and $M^+$ are the subsets of the parameter space of $\boldsymbol{\Omega}$ in which $\boldsymbol{\Omega}$ is a positive definite precision matrix. The solution of \eqref{glasso} is called the graphical LASSO estimator, and there have been many research studies on this model in recent years, including those by Meinshausen and B\"{u}hlmann (2006), Yuan and Lin (2007), Banerjee et al. (2008), Friedman et al. (2008), and Guo et al. (2011) among others.

Note that the penalty in \eqref{glasso} is equivalent to the logarithm of
\begin{align}
\label{lasso.prior}
 p(\omega_{ij}) = \begin{cases}
 \lambda e^{-\lambda\omega_{ii}}, & (i=j); \\
 \frac{\lambda}2 e^{-\lambda|\omega_{ij}|}, &  (i\ne j).
 \end{cases}
\end{align}
From the viewpoint of Bayesian statistics, as in Marlin et al. (2009) and Marlin and Murphy (2009), the graphical LASSO estimator is a maximum a posteriori estimator of $\boldsymbol{\Omega}$ in which the prior distribution of each diagonal element is exponential and that of each off-diagonal element is Laplace as in \eqref{lasso.prior}. This is a natural extension of the original Bayesian LASSO by Park and Casella (2008) who have extended the LASSO regression by Tibshirani (1996) to a Bayesian counterpart.

Based on this interpretation, Wang (2012) and Khondker et al. (2013) have independently proposed Markov chain sampling algorithms to generate the precision matrix $\boldsymbol{\Omega}$ from its posterior distribution. Wang (2012) has developed a Gibbs sampling algorithm, while Khondker et al. (2013) have devised a Metropolis-Hastings algorithm that can generate a positive definite precision matrix. In this paper, we explore Wang's (2012) approach since it is a pure Gibbs sampler and does not suffer from a low acceptance rate even if the dimension of $\boldsymbol{\Omega}$ is high.

Let us briefly review Wang's (2012) algorithm (block Gibbs sampler), which we will discuss in more detail in Section 2. Wang's (2012) block Gibbs sampler generates the $i$-th diagonal element $\omega_{ii}$ and the off-diagonal elements in the $i$-th column (or row) alternatively in the following fashion.\footnote{Although we have simplified the steps here for a brief overview of the algorithm, there are other steps for sampling the shrinkage parameters. Please see Section 2 for details.
}
\begin{itembox}{\emph{Block Gibbs sampler for the precision matrix}}
For $i=1,\dots,p$, repeat \textbf{Step 1} to \textbf{Step 3}.
\begin{description}
\item[Step 1:] Partition $\boldsymbol{\Omega}$ into the $i$-th diagonal element $\omega_{ii}$, the off-diagonal elements $(\omega_{1i},\dots,\omega_{i-1,i},\ \omega_{i+1,i},\dots,\omega_{pi})$, and the rest.
\item[Step 2:] Generate $(\omega_{1i},\dots,\omega_{i-1,i},\ \omega_{i+1,i},\dots,\omega_{pi})$ from the full conditional posterior distribution.
\item[Step 3:] Generate $\omega_{ii}$ from the full conditional posterior distribution.
\end{description}
\end{itembox}
These full conditional posterior distributions will be derived in Section 2. Since Wang's (2012) block Gibbs sampler enables us to generate $\boldsymbol{\Omega}$ from the posterior distribution so easily, it has become an indispensable building block for recent applied research on the Bayesian analysis of Gaussian graphical models. For example, as natural extensions of the block Gibbs sampler, Wang (2015) has extended the original algorithm to a graphical spike-and-slab model, while Li et al. (2019) have applied it to a graphical horseshoe model.

\begin{table}[tb]
  \caption{The number of violations in the positive definiteness of $\boldsymbol{\Omega}$}
  \begin{center}
  \begin{tabular}{rcccccc}\hline
         $p$  & AR(1) & AR(2) & Block & Star & Circle & Full\\ \hline
        30 & 7,644 & 561 & 12 & 27 & 77,768 & 14 \\ 
               & (2.55) & (0.19)&  (0.00)&(0.01) & (25.88) &(0.00)\\
        100 & 566 & 9  & 0 & 2,524 & 205,093 & 0 \\
        & (0.06)&(0.00) & (0.00) &(0.25) & (20.51) &(0.00)\\ \hline
        Notes: & \multicolumn{6}{l}{(a) The number of generated $\boldsymbol{\Omega}$'s is $p\times 10,000$.} \\
        & \multicolumn{6}{l}{(b) The figures in parentheses are the \% ratios.} \\
  \end{tabular}
  \end{center}
\end{table}

Although the block Gibbs sampler and its variants proposed in recent years are nice and elegant, the precision matrix $\boldsymbol{\Omega}$ generated with these sampling algorithms is not necessarily positive definite because the off-diagonal elements of $\boldsymbol{\Omega}$ are not generated from $M^+$ in \textbf{Step 2}. To a varying degree, this problem occurs regardless of whether the choice of the prior distribution is LASSO (Wang [2012]), spike-and-slab prior (Wang [2015]), or horseshoe prior (Li et al. [2019]); although, a strong shrinkage prior may somehow offset the lack of positive definiteness. To demonstrate our point, herein, we run Monte Carlo experiments similar to those conducted by Wang (2012). We generate data sets with six different graph structures (AR(1), AR(2), Block, Star, Circle, and Full) and two different dimensions $(p=30, 100)$, and apply the block Gibbs sampler for the Bayesian adaptive LASSO\footnote{We  have explained the Bayesian adaptive LASSO in Section 2. In our experience, violation of the positive definiteness occurs whether it is adaptive or not.} in which the shrinkage parameter $\lambda$ may differ from element to element in $\boldsymbol{\Omega}$. The number of iterations in the block Gibbs sampler is 10,000 for each experiment. Thus, if we count every $\boldsymbol{\Omega}$ that is partially updated from \textbf{Step 1} to \textbf{Step 3} as distinctive, we have 300,000 $(p=30)$ or 10,000,000 $(p=100)$ replications of $\boldsymbol{\Omega}$ in one experiment. The results of the Monte Carlo experiments are summarized in Table 1. In the case of $p = 30$, violation of the positive definiteness occurs in all designs. In particular, about one quarter of the generated $\boldsymbol{\Omega}$'s do not satisfy the positive definiteness in the Circle design. In the case of $p =100$, the violation of the positive definiteness is less severe for some designs, but the ratio of violation is still high (20.51\%) in the Circle design.

To address this issue, we propose improving Wang's (2012) block Gibbs sampler so that the generated $\boldsymbol{\Omega}$ will never fail to be positive definite. Although it seems too intractable to guarantee the positive definiteness of $\boldsymbol{\Omega}$ in each cycle of the block Gibbs sampler, the hit-and-run algorithm by B\'{e}lisle et al. (1993) is applicable to the Bayesian (adaptive) graphical LASSO in a fairly straightforward manner, and the resultant algorithm is a pure Gibbs sampler without the Metropolis-Hastings step. Therefore, our proposed algorithm enjoys the same efficiency as Wang's (2012) but can prevent $\boldsymbol{\Omega}$ from violating the positive definiteness.

The main body of this paper is organized as follows: In Section 2, we briefly review Wang's (2012) block Gibbs sampling algorithm for the Bayesian adaptive graphical LASSO, though Wang (2012) has also derived an algorithm for the Bayesian graphical LASSO with the common shrinkage parameter. This is because the core part of the block Gibbs sampling algorithm is almost identical in both prior settings. In Section 3, we discuss why the positive definiteness of the precision matrix is violated in Wang's (2012) algorithm and derive a modified Gibbs sampling algorithm that guarantees positive definiteness. In Section 4, we compare our proposed algorithm with Wang's (2012) in several Monte Carlo experiments and report the results of the performance comparison. Finally, in Section 5, we state our concluding remarks.

\section{Review of Wang's (2012) Algorithm}
In this section, we briefly review a Gibbs sampling algorithm developed by Wang (2012). Although, Wang (2012) derived it for the Bayesian graphical LASSO with the prior distribution \eqref{lasso.prior}, we consider a more general prior setting that allows $\lambda$ in \eqref{lasso.prior} to vary for each element of precision matrix $\boldsymbol{\Omega}$, namely 
\begin{align}
\label{ada.prior}
 p(\omega_{ij}) = \begin{cases}
 \lambda_{ii} e^{-\lambda_{ii}\omega_{ii}}, & (i=j); \\
 \frac{\lambda_{ij}}2 e^{-\lambda_{ij}|\omega_{ij}|}, &  (i\ne j),
 \end{cases}
\end{align}
which is called the adaptive graphical LASSO. Since Wang (2012) demonstrated that the Bayesian adaptive LASSO outperforms its nonadaptive counterpart in terms of parameter estimation and graphical structure learning, we will illustrate the Gibbs sampling algorithm for the adaptive LASSO in detail.

 To derive the Gibbs sampling algorithm, Wang (2012) utilized the well-known fact that the Laplace distribution in \eqref{ada.prior} is expressed as a scale mixture of normal distributions with the exponential distribution:
\begin{equation}
\label{scale.mixture}
 \omega_{ij}|\tau_{ij}\sim\mathcal{N}(0,\tau_{ij}),\quad
 \tau_{ij}\sim \mathcal{E}xp\left(\frac{\lambda_{ij}^2}2\right).
\end{equation}
By using gamma distribution $\mathcal{G}a(r,s)$ as the common prior for $\lambda_{ij}$ $(1\leqq i \leqq j\leqq p)$, we obtain the joint posterior distribution\footnote{Wang (2012) assumed that the prior distribution of each diagonal element $\omega_{ii}$ is $\frac{\lambda_{ii}}{2}\exp\left(-\frac{\lambda_{ii}}{2}\omega_{ii}\right)$ instead of $\lambda_{ii}\exp\left(-\lambda_{ii}\omega_{ii}\right)$. This is because Wang (2012) employed $\Vert\boldsymbol{\Omega}\Vert_1 = \sum_{i=1}^p\sum_{j=1}^p|\omega_{ij}|$ as the penalty, in which each off-diagonal element $\omega_{ij}$ $(i\ne j)$ appears twice. However, ours is $\Vert\boldsymbol{\Omega}\Vert_1 = \sum_{i=1}^p\sum_{j=1}^i|\omega_{ij}|$, which includes the lower triangular part of $\boldsymbol{\Omega}$ only.} of $\boldsymbol{\omega}=\{\omega_{ij}\}_{i\leqq j}$, $\boldsymbol{\tau}=\{\tau_{ij}\}_{i<j}$ and $\boldsymbol{\lambda}=\{\lambda_{ij}\}_{i\leqq j}$ as
\begin{align}
\label{ada.posterior}
p(\boldsymbol{\omega},\boldsymbol{\tau},\boldsymbol{\lambda}|\boldsymbol{Y}) &\propto |\boldsymbol{\Omega}|^{\frac{n}2}\exp\left[-\frac12\mathrm{tr}(\boldsymbol{S}\boldsymbol{\Omega})\right] \prod_{i=1}^{p}\lambda_{ii}e^{-\lambda_{ii}\omega_{ii}} \nonumber \\
 &\quad \times \prod_{i<j}\frac1{\sqrt{2\pi\tau_{ij}}}\exp\left(-\frac{\omega_{ij}^2}{2\tau_{ij}}\right) \frac{\lambda_{ij}^2}2\exp\left(-\frac{\lambda_{ij}^2}2\tau_{ij}\right)\mathbf{1}_{M^+}(\boldsymbol{\Omega}) \nonumber \\
 &\quad \times \prod_{i\leqq j}\lambda_{ij}^{r-1}e^{-s\lambda_{ij}},
\end{align}
where $\mathbf{1}_{M^+}(\boldsymbol{\Omega})$ is the indicator function that will be equal to 1 if $\boldsymbol{\Omega}\in M^+$; otherwise, it is equal to 0. To construct a Gibbs sampler for the posterior distribution in \eqref{ada.posterior}, we need to derive all full conditional posterior distributions for $\boldsymbol{\omega}$, $\boldsymbol{\tau}$, and $\boldsymbol{\lambda}$.

It is straightforward to show that the full conditional posterior distribution of $1/\tau_{ij}$ $(1\leqq i < j \leqq p)$ is the inverse Gaussian distribution:
\begin{equation}
\label{fc.tau}
\left.\frac1{\tau_{ij}}\right|\boldsymbol{\theta}_{-\tau_{ij}}, \boldsymbol{Y} \sim \mathcal{IG}\left(\frac{\lambda_{ij}}{|\omega_{ij}|},\lambda^2_{ij}\right),
\end{equation}
while that of $\lambda_{ij}$ $(1\leqq i \leqq j \leqq p)$ is the gamma distribution:
\begin{equation}
\label{fc.lambda}
\lambda_{ij} |\boldsymbol{\theta}_{-\lambda_{ij}}, \boldsymbol{Y} \sim \mathcal{G}a \left(r + 1, s + |\omega_{ij}|\right),
\end{equation}
where $\boldsymbol{\theta}$ represents the vector of all parameters and latent variables in the model and expressions such as $\boldsymbol{\theta}_{-x}$ indicate that a parameter $x$ is excluded from $\boldsymbol{\theta}$. Note that $\tau_{ij}$ is integrated out in \eqref{fc.lambda}.

To generate $\boldsymbol{\omega}$ from the full conditional posterior distribution, Wang (2012) proposed a Gibbs sampling algorithm that iteratively generates each diagonal element and the corresponding off-diagonal elements of the precision matrix $\boldsymbol{\Omega}$ from their full conditional posterior distributions, i.e., the block Gibbs sampler. The block Gibbs sampler is based on the following partition of $\boldsymbol{\Omega}$:
\begin{equation}
\label{partition1}
\boldsymbol{\Omega} =
\begin{bmatrix}
\boldsymbol{\Omega}_{11} & \boldsymbol{\omega}_{12} \\
\boldsymbol{\omega}_{12}^{\intercal} & \omega_{22} 
\end{bmatrix},
\end{equation}
where $\boldsymbol{\Omega}_{11}$ is a $(p-1\times p-1)$ matrix, $\boldsymbol{\omega}_{12}$ is a $(p-1\times 1)$ vector, and $\omega_{22}$ is a scalar. Without a loss of generality we can rearrange the rows and columns of $\boldsymbol{\Omega}$, so that the lower-right corner of $\boldsymbol{\Omega}$, $\omega_{22}$ is the diagonal element to be generated from its full conditional posterior distribution. Likewise, we can partition $\boldsymbol{S}$, $\boldsymbol{\Upsilon}$, and $\boldsymbol{\lambda}$ as
\begin{equation}
\label{partition2}
 \boldsymbol{S} =
 \begin{bmatrix}
\boldsymbol{S}_{11} & \boldsymbol{s}_{12} \\
\boldsymbol{s}_{12}^{\intercal} & s_{22} 
\end{bmatrix},\quad
 \boldsymbol{\Upsilon} =
 \begin{bmatrix}
\boldsymbol{\Upsilon}_{11}  & \boldsymbol{\tau}_{12} \\
\boldsymbol{\tau}_{12}^{\intercal} & 0
\end{bmatrix},\quad
 \boldsymbol{\lambda} =
 \begin{bmatrix}
\boldsymbol{\lambda}_{12}  \\ \lambda_{22} \\
\end{bmatrix},
\end{equation}
where $\boldsymbol{\Upsilon}$ is a $(p\times p)$ symmetric matrix in which the off-diagonal $(i,j)$ element is $\tau_{ij}$ and all diagonal elements are equal to zero, while $\lambda_{22}$ is the element in $\boldsymbol{\lambda}$ that corresponds with the diagonal element $\omega_{22}$ in the prior distribution \eqref{ada.prior}.

With the partition of $\boldsymbol{\Omega}$ in \eqref{partition1} and $\boldsymbol{S}$ in \eqref{partition2}, we have
\begin{align*}
 \mathrm{tr}\left(\boldsymbol{S}\boldsymbol{\Omega}\right)	
 &= s_{22}\omega_{22}
 + 2\boldsymbol{s}_{12}^{\intercal}\boldsymbol{\omega}_{12}
 + \mathrm{tr}\left(\boldsymbol{S}_{11}\boldsymbol{\Omega}_{11}\right),
\end{align*}
and
\begin{align*}
\left|\boldsymbol{\Omega}\right|
 &= \left|\omega_{22}-\boldsymbol{\omega}_{12}^{\intercal}\boldsymbol{\Omega}_{11}^{-1}\boldsymbol{\omega}_{12}\right|\left|\boldsymbol{\Omega}_{11}\right|.
 \end{align*}
Then, the likelihood can be expressed as 
\begin{align}
\label{likelihood.trace.form2}
 p(\boldsymbol{Y}|\boldsymbol{\Omega})
 &\propto |\boldsymbol{\Omega}|^{\frac{n}2}\exp\left[-\frac12\mathrm{tr}(\boldsymbol{S}\boldsymbol{\Omega})\right] \nonumber \\
 &\propto
 \left|\omega_{22}-\boldsymbol{\omega}_{12}^{\intercal}\boldsymbol{\Omega}_{11}^{-1}\boldsymbol{\omega}_{12}\right|^{\frac{n}2}
 \left|\boldsymbol{\Omega}_{11}\right|^{\frac{n}2} \nonumber \\
 &\quad \times\exp\left[-\frac12\left\{s_{22}\omega_{22}
 + 2\boldsymbol{s}_{12}^{\intercal}\boldsymbol{\omega}_{12}
 + \mathrm{tr}\left(\boldsymbol{S}_{11}\boldsymbol{\Omega}_{11}\right)
 \right\}\right].
\end{align}
Wang (2012) reparametrized $(\omega_{22}, \boldsymbol{\omega}_{12})$ to $(\gamma, \boldsymbol{\beta})$, where 
\begin{equation}
\label{transformation}	
\gamma = \omega_{22}-\boldsymbol{\omega}_{12}^{\intercal}\boldsymbol{\Omega}_{11}^{-1}\boldsymbol{\omega}_{12}, \quad \boldsymbol{\beta} = \boldsymbol{\omega_{12}}.
\end{equation}
Thus, the likelihood \eqref{likelihood.trace.form2} can be expressed as follows:
\begin{align}
\label{likelihood.trace.form3}
 p(\boldsymbol{Y}|\boldsymbol{\Omega})
 & \propto \gamma^{\frac{N}2} \exp\left[-\frac12 \left\{s_{22}\gamma 
 + s_{22}\boldsymbol{\beta}^{\intercal}\boldsymbol{\Omega}_{11}^{-1}\boldsymbol{\beta} 
 + 2s_{12}^{\intercal}\boldsymbol{\beta} 
 + \mathrm{tr}(\boldsymbol{S}_{11}\boldsymbol{\Omega}_{11})\right\}\right] \nonumber \\
 & \propto
 \gamma^{\frac{N}2}
 \exp\biggl[-\frac12\Bigl\{s_{22}\gamma
 + s_{22}\boldsymbol{\beta}^{\intercal}\boldsymbol{\Omega}_{11}^{-1}\boldsymbol{\beta} 
 + 2s_{22}\boldsymbol{\beta}\Bigr\}\biggr].
\end{align}
With the adaptive prior \eqref{ada.prior} and the flat prior
$p(\gamma)\propto\text{constant}$, Wang (2012) proposed using
\begin{align}
\label{fc.beta}
 & \boldsymbol{\beta}|\boldsymbol{\theta}_{-\boldsymbol{\beta}},\boldsymbol{Y}
 \sim \mathcal{N}\left(-\boldsymbol{C}\boldsymbol{s}_{12},\ \boldsymbol{C}\right), \\ \nonumber
 & \boldsymbol{C} = \left\{(s_{22} + 2\lambda_{22})\boldsymbol{\Omega}_{11}^{-1}
 + \boldsymbol{D}_{\boldsymbol{\tau}}^{-1}\right\}^{-1},\quad
 \boldsymbol{D}_{\boldsymbol{\tau}} = \mathrm{diag}(\boldsymbol{\tau}_{12}), \\ 
 \label{fc.gamma}
 & \gamma|\boldsymbol{\theta}_{-\gamma},\boldsymbol{Y}
 \sim \mathcal{G}a\left(\frac{n}2+1,\ \frac{s_{22}}{2}+\lambda_{22}\right) 
\end{align}
as the full conditional posterior distribution of $\gamma$ and $\boldsymbol{\beta}$.

In summary, Wang's (2012) block Gibbs sampler is given as follows:\footnote{Since this algorithm is a Gibbs sampler, there should be no problem in calculating the posterior distribution even if the order of the steps is changed. In fact, Wang (2012) sampled \emph{Step 4} and \emph{Step 5} first in the code disclosed before. However, we confirmed that Wang's code disclosed before does not work if we swap \emph{Step 2} and \emph{Step 3}. This implies that Wang's algorithm cannot sample from the correct posterior distribution. In contrast, we confirmed that our algorithm proposed in Section 3 works even if we exchange \emph{Step 2} and \emph{Step 3}.},\footnote{In Wang's (2012) study, \emph{Step 4} and \emph{Step 5} are calculated together outside the for loop, but since there is no essential difference, they are shown in the for loop here.
}
\begin{itembox}{\emph{Block Gibbs sampler for all parameters}}
For $i=1,\dots,p$, repeat \textbf{Step 1} to \textbf{Step 5}.
\begin{description}
\item[\it Step 1:] Rearrange $\boldsymbol{\Omega}$, $\boldsymbol{S}$, $\boldsymbol{\Upsilon}$, and $\boldsymbol{\lambda}$ so that $\omega_{ii}$ is in the place of $\omega_{22}$ in $\boldsymbol{\Omega}$ and partition them as in \eqref{partition1} and \eqref{partition2}.
\item[\it Step 2:] If $i \geqq 2$, $\boldsymbol{\beta}\leftarrow\mathcal{N}\left(-\boldsymbol{C}\boldsymbol{s}_{12},\boldsymbol{C}\right)$, and set $\boldsymbol{\omega}_{12}=\boldsymbol{\beta}$.
\item[\it Step 3:] $\gamma\leftarrow\mathcal{G}a\left(\frac{n}2+1,\frac{s_{22}}{2}+\lambda_{22} \right)$, and set $\omega_{22}=\gamma+\boldsymbol{\omega}_{12}\boldsymbol{\Omega}_{11}^{-1}\boldsymbol{\omega}_{12}$.
\item[\it Step 4:] $\lambda_{12}\leftarrow\mathcal{G}a \left(r + 1, s + |\boldsymbol{\omega}_{12}|\right)$.
\item[\it Step 5:] $\upsilon\leftarrow\mathcal{IG}\left(\frac{\lambda_{12}}{|\boldsymbol{\omega}_{12}|},\lambda^2_{12}\right)$, and set $\tau_{12} = 1/\upsilon$.
\end{description}
\end{itembox}

\section{Proposed Algorithm}
As we pointed out in the introduction, Wang's (2012) block Gibbs sampler does not necessarily guarantee the positive definiteness of the generated $\boldsymbol{\Omega}$'s. Therefore, in this section, we propose an efficient sampling method to generate them under the positive definiteness constraint: $\boldsymbol{\Omega}\in M^+$.

First, let us derive the full conditional posterior distribution of $\gamma$. Here, we need to take care in choosing the prior distribution of $(\gamma,\beta)$. Given that $\Omega$ from the previous iteration of the block Gibbs sampler is positive definite, the newly generated $\omega_{22}$ and $\boldsymbol{\omega}_{12}$ must satisfy
\begin{equation}
\label{zyoken}
 \omega_{22} > \boldsymbol{\omega}_{12}^{\intercal}\boldsymbol{\Omega}_{11}^{-1}\boldsymbol{\omega}_{12}
\end{equation}
to ensure that the updated $\boldsymbol{\Omega}$ is also positive definite. This condition \eqref{zyoken} requires
\[
    \gamma =  \omega_{22} - \boldsymbol{\beta}^{\intercal}\boldsymbol{\Omega}_{11}^{-1}\boldsymbol{\beta} > 0.
\]
as the prior distribution of $\gamma$. In other words, the conditional prior distribution of $\gamma$ given $\boldsymbol{\beta}$ and $\boldsymbol{\Omega}_{11}$ must be
\begin{equation}
\label{omega22.prior}
 p(\gamma|\boldsymbol{\beta}, \boldsymbol{\Omega}_{11})
 \propto \lambda_{22}\exp\left(-\lambda_{22}\gamma \right)\mathbf{1}_{M_{\gamma}^+}(\gamma),
\end{equation}
where $M_{\gamma}^+ = \{\gamma: \gamma > 0\}$. 
Therefore, by ignoring the parts that do not depend on $\gamma$ in \eqref{likelihood.trace.form2}, we obtain
\begin{align}
\label{fc.omega22.pdf}
 & p(\gamma|\boldsymbol{\theta}_{-\gamma},\boldsymbol{Y}) \nonumber \\
 &\quad \propto
 \left|\gamma\right|^{\frac{n}2}
 \exp\left(-\frac{s_{22}}2\gamma\right) \times \exp\left(-\lambda_{22}\gamma\right)\mathbf{1}_{M_{\gamma}^+}(\gamma) \nonumber\\
 &\quad \propto
 \left|\gamma\right|^{\frac{n}2}
 \exp\left[-\frac{s_{22}+2\lambda_{22}}2\left(\gamma\right)\right] \mathbf{1}_{M_{\gamma}^+}(\gamma).
\end{align}
The full conditional posterior distribution of $\gamma$ in \eqref{fc.omega22.pdf} is the gamma distribution:
\begin{equation}
\label{fc.omega22}
 \gamma|\boldsymbol{\theta}_{-\gamma},\boldsymbol{Y}
 \sim \mathcal{G}a\left(\frac{n}2+1,\ \frac{s_{22}}{2}+\lambda_{22}\right)
\end{equation}
Obviously, the distribution of $\gamma$ in \eqref{fc.omega22} is equivalent to that in \eqref{fc.gamma}. Thus, \eqref{fc.omega22} and \eqref{fc.gamma} are basically identical to each other, and $\gamma$ generated from either \eqref{fc.omega22} or \eqref{fc.gamma} always satisfies the positive definiteness condition \eqref{zyoken} because random variables generated from the gamma distribution always have positive values.

Next, let us derive the full conditional posterior distribution of $\boldsymbol{\beta}$. For the same reason as in \eqref{omega22.prior}, the conditional prior distribution of $\boldsymbol{\beta}$ must be the following truncated multivariate normal distribution:
\begin{equation}
\label{omega12.prior}
 p(\boldsymbol{\beta}|\gamma, \boldsymbol{\Omega}_{11})
 \propto \exp\left(-\frac12\boldsymbol{\beta}^{\intercal}\boldsymbol{D}_{\boldsymbol{\tau}}^{-1}\boldsymbol{\beta}\right)\mathbf{1}_{M_{\beta}^+}(\boldsymbol{\beta}),
\end{equation}
where $M_{\beta}^+ = \{\boldsymbol{\beta}: \omega_{22} > \boldsymbol{\beta}^{\intercal}\boldsymbol{\Omega}_{11}^{-1}\boldsymbol{\beta}\}$. As a result, the full conditional posterior distribution of $\boldsymbol{\beta}$ is also a truncated multivariate normal distribution:
\begin{equation}
\label{fc.omega12}
 \boldsymbol{\beta}|\boldsymbol{\theta}_{-\boldsymbol{\beta}}, \boldsymbol{Y}
 \sim \mathcal{N}\left(-\boldsymbol{C}\boldsymbol{s}_{12},\ \boldsymbol{C}\right)\mathbf{1}_{M_{\beta}^+}(\boldsymbol{\beta}).
\end{equation}
However, Wang (2012) proposed using the unconstrained multivariate normal distribution \eqref{fc.beta}, which does not impose the truncation $\mathbf{1}_{M_{\beta}^+}(\boldsymbol{\beta})$, to generate $\boldsymbol{\beta}$. 
Consequently, if we generate $\boldsymbol{\beta}$ from \eqref{fc.beta}, there is no guarantee that the newly updated $\boldsymbol{\omega}_{12}$ will satisfy the positive definiteness condition \eqref{zyoken}. This is why generated $\boldsymbol{\Omega}$'s are not always positive definite, as shown in Table 1. Therefore, to ensure the positive definiteness of $\boldsymbol{\Omega}$, it is preferable to use the truncated multivariate normal distribution \eqref{fc.omega12} in the block Gibbs sampler.

Since both the naive rejection method and Metropolis-Hastings algorithm are inefficient, even for a modest-size graphical model, we can apply the hit-and-run algorithm (B\'{e}lisle et al. [1993]) to generate $\boldsymbol{\beta}$ from the truncated multivariate normal distribution \eqref{fc.omega22}.
\begin{itembox}{\emph{Hit-and-run algorithm}}
\begin{description}
 \item[\it Step 1:] Pick a point $\boldsymbol{\alpha}$ on the unit sphere randomly as $\boldsymbol{\alpha}=\frac{\boldsymbol{z}}{\Vert\boldsymbol{z}\Vert}$, $\boldsymbol{z}\sim\mathcal{N}(\boldsymbol{0},\boldsymbol{I})$.
 \item[\it Step 2:] Generate a random scalar $\kappa$ from the distribution with the density
\begin{equation}
\label{fc.kappa}
 f(\kappa) \propto p(\boldsymbol{\beta}+\kappa\boldsymbol{\alpha})
 \mathbf{1}_{M_{\beta}^+}(\boldsymbol{\beta}+\kappa\boldsymbol{\alpha}),
\end{equation}
where $p(\cdot)$ is the density of $\mathcal{N}\left(-\boldsymbol{C}\boldsymbol{s}_{12}, \boldsymbol{C}\right)$ in \eqref{fc.omega12}.
 \item[\it Step 3:] Set $\boldsymbol{\beta}+\kappa\boldsymbol{\alpha}$ as the new $\boldsymbol{\beta}$.
\end{description}
\end{itembox}
It is straightforward to show that the distribution of $\kappa$ in \eqref{fc.kappa} is
\begin{equation}
\label{fc.kappa2}
 \kappa \sim \mathcal{N}\left(\mu_{\kappa},\sigma_{\kappa}^2\right)\mathbf{1}_{M_{\beta}^+}(\boldsymbol{\beta}+\kappa\boldsymbol{\alpha}),
\end{equation}
where 
\[
 \mu_{\kappa} = -\frac{\boldsymbol{s}_{12}^{\intercal}\boldsymbol{\alpha}+\boldsymbol{\beta}^{\intercal}\boldsymbol{C}^{-1}\boldsymbol{\alpha}}
 {\boldsymbol{\alpha}^{\intercal}\boldsymbol{C}^{-1}\boldsymbol{\alpha}},\quad 
 \sigma_{\kappa}^2 = \frac1{\boldsymbol{\alpha}^{\intercal}\boldsymbol{C}^{-1}\boldsymbol{\alpha}}.
\]
The indicator function $\mathbf{1}_{M_{\beta}^+}(\boldsymbol{\beta}+\kappa\boldsymbol{\alpha})$ is equal to 1 if and only if
\[
 (\boldsymbol{\beta}+\kappa\boldsymbol{\alpha})^{\intercal}
 \boldsymbol{\Omega}_{11}^{-1}(\boldsymbol{\beta}+\kappa\boldsymbol{\alpha}) - (\gamma + \boldsymbol{\beta}^{\intercal}\boldsymbol{\Omega}_{11}^{-1}\boldsymbol{\beta}) < 0.
\]
This means that $\kappa$ must satisfy
\[
 \underbrace{\left(\boldsymbol{\alpha}^{\intercal}\boldsymbol{\Omega}_{11}^{-1}\boldsymbol{\alpha}\right)}_{\displaystyle a}\kappa^2
 + 2\underbrace{\left(\boldsymbol{\beta}^{\intercal}\boldsymbol{\Omega}_{11}^{-1}\boldsymbol{\alpha}\right)}_{\displaystyle b}\kappa
 + \underbrace{(-\gamma)}_{\displaystyle c} < 0.
\]
Note that $a > 0$, $c < 0$ as long as the current $\boldsymbol{\Omega}$ is positive definite, which implies that the quadratic equation $a\kappa^2 + 2b\kappa + c = 0$ has two distinctive real roots. Therefore, the distribution in \eqref{fc.kappa2} is the truncated univariate normal distribution on the interval:
\[
 R^{+} = \left\{\kappa : \frac{-b-\sqrt{b^2-ac}}{a} < \kappa <  \frac{-b+\sqrt{b^2-ac}}{a}\right\}.
\]
Thus, using the hit-and-run algorithm, sampling from the seemingly intractable distribution \eqref{fc.omega22} is reduced to sampling from the truncated univariate normal distribution:
\[
 \kappa \sim \mathcal{N}\left(\mu_{\kappa},\sigma_{\kappa}^2\right)\mathbf{1}_{R^+}(\kappa),
\]
and the sampling procedure becomes much simpler.

By replacing \eqref{fc.gamma} in \textbf{Step 2} with \eqref{fc.omega22} and \eqref{fc.beta} in \textbf{Step 3} with the hit-and-run algorithm, we obtain the modified block Gibbs sampler as follows:

\begin{itembox}{\emph{Modified block Gibbs sampler}}
For $i=1,\dots,p$, repeat \textbf{Step 1} to \textbf{Step 5}.
\begin{description}
\item[\it Step 1:] Rearrange $\boldsymbol{\Omega}$, $\boldsymbol{S}$, $\boldsymbol{\Upsilon}$, and $\boldsymbol{\lambda}$ so that $\omega_{ii}$ is in the place of $\omega_{22}$ in $\boldsymbol{\Omega}$ and partition them as in \eqref{partition1} and \eqref{partition2}.
\item[\it Step 2:] If $i \geqq 2$,
\begin{enumerate}
\item[(a)] $\boldsymbol{z}\leftarrow\mathcal{N}(\boldsymbol{0},\boldsymbol{I})$, and set $\boldsymbol{\alpha}=\frac{\boldsymbol{z}}{\Vert\boldsymbol{z}\Vert}$.
\item[(b)] $\kappa\leftarrow\mathcal{N}\left(\mu_{\kappa},\sigma_{\kappa}^2\right)\mathbf{1}_{R^{+}}(\kappa)$, and update the old $\boldsymbol{\beta}$ with $\boldsymbol{\beta}+\kappa\boldsymbol{\alpha}$. Then, set $\boldsymbol{\omega}_{12} = \beta$.
\end{enumerate}
\item[\it Step 3:] $\gamma \leftarrow\mathcal{G}a\left(\frac{n}2+1,\frac{s_{22}}{2}+\lambda_{22} \right)$ and set $\omega_{22}=\gamma+\boldsymbol{\omega}_{12}\boldsymbol{\Omega}_{11}^{-1}\boldsymbol{\omega}_{12}$.
\item[\it Step 4:] $\lambda_{12}\leftarrow\mathcal{G}a \left(r + 1, s + |\boldsymbol{\omega}_{12}|\right)$.
\item[\it Step 5:] $\upsilon\leftarrow\mathcal{IG}\left(\frac{\lambda_{12}}{|\boldsymbol{\omega}_{12}|},\lambda^2_{12}\right)$, and set $\tau_{12} = 1/\upsilon$.
\end{description}
\end{itembox}

\section{Performance Comparison}
\subsection{Simulation Study}
In this section, we report the results of the Monte Carlo experiments to compare our modified block Gibbs sampler with Wang's (2012) original algorithm in terms of accuracy in the parameter estimation and graphical structure learning. For brevity, we shall refer to Wang's (2012) original algorithm as the BGS (block Gibbs sampler) and our modified version as the HRS (hit-and-run sampler). Following Wang (2012), we examined the following six different specifications of the Gaussian graphical model in the Monte Carlo experiments:
\begin{enumerate}
\item[(a)] AR(1): $\sigma_{ij} = 0.7^{|i-j|}$.
\item[(b)] AR(2): $\omega_{ii}=1.0$, $\omega_{i,i-1}=\omega_{i-1,i}=0.5$, and $\omega_{i,i-2}=\omega_{i-2,i}=0.25$.
\item[(c)] Block: $\sigma_{ii}=1$, $\sigma_{ij}=0.5$ for $1\leq i\neq j \leq p/2$ , $\sigma_{ij}=0.5$ for $p/2+1\leq i \neq j \leq 10$, and $\sigma_{ij}=0.0$ otherwise.
\item[(d)] Star: $\omega_{ii}=1.0$, $\omega_{1,i}=\omega_{i,1}=0.1$, and $\omega_{ij}=0.0$ otherwise.
\item[(e)] Circle: $\omega_{ii} = 2.0$, $\omega_{i-1,i} = \omega_{i, i-1} = 1.0$, $\omega_{1p}=\omega_{p1}=0.9$.
\item[(f)] Full: $\omega_{ii}=2.0$, $\omega_{ij}=1.0$ for $i\neq j$.
\end{enumerate}
Here, $\sigma_{ij}$ $(1\leqq i,\ j \leqq p)$ is the $(i,j)$ element of the covariance matrix $\boldsymbol{\Omega}^{-1}$ in the Gaussian graphical model.

The other settings for the Monte Carlo experiments also mirrored Wang's (2012). For each model, we generated a sample of $(p\times 1)$ random vectors $\boldsymbol{y}_1,\dots,\boldsymbol{y}_n$ independently from $\mathcal{N}(\boldsymbol{0},\boldsymbol{\Omega}^{-1})$. We considered two cases: $(n, p) = (50, 30)$ and $(n, p) = (200, 100)$. Thus, we tried $12\ (=6\times 2)$ scenarios in the experiments. The hyperparameters in the prior distribution of $\lambda_{ij}$ were $r = 10^{-2}$ and $s = 10^{-6}$. For both the BGS and HRS, the number of burn-in iterations were 5,000, and the Monte Carlo sample from the following 10,000 iterations was used in the Bayesian inference.\footnote{The same simulation design (specifications of $\boldsymbol{\Omega}$, combinations of $(n,p)$, hyperparameters, burn-in iterations, and the size of the Monte Carlo sample) was used in producing the results in Table 1.} We repeated each simulation scenario 50 times and obtained a set of point estimates of $\boldsymbol{\Omega}$. All computations were implemented on a workstation with 64 GB RAM and a six-core 3.4 GHz Intel Xeon processor using Python 3.6.1. For the BGS, we rewrote Wang's disclosed MATLAB code "BayesGLassoGDP.m" into Python and used it. Although not mentioned in Wang's (2012) study, there is a part that arbitrarily cuts a range of random number generations of $\lambda_{12}$ and $\tau_{12}$ in Wang's disclosed code. We took over this adjustment in our rewritten Python code because the BGS calculation resulted in an error if we excluded the adjustment. The HRS required additional computations because it explicitly imposed the positive definite constraint $\boldsymbol{\Omega}\in M^+$, but we observed only a modest difference in computation time between the HRS and BGS.

To compare the HRS with the BGS in terms of accuracy in the point estimation of the precision matrix $\boldsymbol{\Omega}$, we computed two sample loss functions, Stein's loss and the Frobenius norm, as measurements of discrepancy between the point estimate and the true $\boldsymbol{\Omega}$. Table 2 shows the sample median loss (Stein's loss in the upper half, and the Frobenius norm in the lower half) of 50 replications in 12 scenarios for the BGS and HRS. The figures in parentheses are the standard errors. The loss was unanimously and substantially smaller in the HRS than in the BGS. This observation was valid not only for the Circle model, in which the positive definiteness of $\boldsymbol{\Omega}$ was most frequently violated as shown in Table 1, but also for the other models with different graphical structures. Interestingly, the HRS outperformed the BGS even for the Full model in which $\boldsymbol{\Omega}$ was not sparse and the estimation loss of the graphical LASSO was expected to be much worse. Furthermore, this tendency was unchanged in either the small $(p = 30)$ or large $(p = 100)$ model. All in all, the results in Table 2 suggest that imposing the positive definiteness constraint remarkably improved the accuracy in the point estimation of $\boldsymbol{\Omega}$ in the Bayesian adaptive graphical LASSO.

\begin{table}[htbp]
\caption{Sample median loss in the point estimation of $\boldsymbol{\Omega}$}
\begin{center}
\begin{footnotesize}
\begin{tabular}{lcccccc}\hline
    & AR(1) & AR(2)& Block & Star & Circle & Full \\ \hline
    & & & & & & \\
\multicolumn{7}{l}{\underline{Stein's loss}} \\
    & \multicolumn{6}{c}{$p = 30$} \\
    & & & & & & \\
BGS & 1.88 & 4.48 & 1.38 & 1.52 & 1.81 & 19.31 \\
    & (0.32) & (0.49) & (0.28) & (0.26) & (0.32) & (0.87) \\ 
HRS & \textbf{0.60} & \textbf{0.76} & \textbf{0.65} & \textbf{0.88} & \textbf{0.55} & \textbf{13.73} \\
    & (0.20) & (0.18) & (0.18) & (0.20) & (0.16) & (0.52) \\ 
    & & & & & & \\
    & \multicolumn{6}{c}{$p = 100$} \\
    & & & & & & \\
BGS & 3.02 & 4.25 & 2.81 & 3.75 & 3.08  & 69.65 \\ 
    & (0.20) & (0.26) & (0.18) & (0.22) & (0.18) & (1.10) \\ 
HRS & \textbf{0.50} & \textbf{0.54} & \textbf{0.51} & \textbf{0.91} & \textbf{0.46} & \textbf{42.17} \\
    & (0.08) & (0.08) & (0.07) & (0.08) & (0.05) & (0.72) \\ \hline
    & & & & & & \\
\multicolumn{7}{l}{\underline{Frobenius norm}} \\
    & \multicolumn{6}{c}{$p = 30$} \\
    & & & & & & \\
BGS & 4.04 & 3.01 & 2.19 & 2.19 & 2.51 & 29.61 \\
    & (0.55) & (0.18) & (0.35) & (0.30) & (0.42) & (0.06) \\
HRS & \textbf{1.53} & \textbf{0.80} & \textbf{1.22} & \textbf{1.33} & \textbf{0.39} & \textbf{19.94} \\
    & (0.27) & (0.13) & (0.22) & (0.22) & (0.07) & (0.49) \\
    & & & & & & \\
    & \multicolumn{6}{c}{$p = 100$} \\
    & & & & & & \\
BGS & 4.38 & 2.33 & 2.90 & 3.20 & 2.59 & 99.61 \\
    & (0.29) & (0.12) & (0.13) & (0.13) & (0.25) & (0.02) \\
HRS & \textbf{1.32} & \textbf{0.60} & \textbf{1.04} & \textbf{1.03} & \textbf{0.25} & \textbf{47.76} \\ 
    & (0.11) & (0.05) & (0.08) & (0.07) & (0.03) & (0.41) \\ \hline
Notes: & \multicolumn{6}{l}{(a) The smaller losses are boldfaced.} \\
        & \multicolumn{6}{l}{(b) The figures in parentheses are the standard errors.} \\
\end{tabular}
\end{footnotesize}
\end{center}
\end{table}

\begin{table}[htbp]
\caption{Accuracy in graphical structure learning}
\begin{center}
\begin{footnotesize}
\begin{tabular}{lrrrrr}\hline
    & AR(1) & AR(2)& Block & Star & Circle \\ \hline
    & & & & & \\
\multicolumn{6}{l}{\underline{Specificity}} \\
    & & & & & \\
    & \multicolumn{5}{c}{$p = 30$} \\
    & & & & & \\
BGS & 6.00 & 10.22 & 7.10 & 6.77 & 12.34 \\
HRS & \textbf{74.82} & \textbf{69.91} & \textbf{80.32} & \textbf{81.08} & \textbf{84.63} \\
    & & & & & \\
    & \multicolumn{5}{c}{$p = 100$} \\
    & & & & & \\
BGS & 10.69 & 20.39 & 12.93 & 12.15 & 28.45 \\
HRS & \textbf{92.70} & \textbf{92.75} & \textbf{94.25} & \textbf{95.22}& \textbf{98.46} \\  \hline
    & & & & & \\
\multicolumn{6}{l}{\underline{Sensitivity}} \\
    & & & & & \\
    & \multicolumn{5}{c}{$p = 30$} \\
    & & & & & \\
BGS & 100.00 & 99.31 & 100.00 & \textbf{96.14} & 100.00 \\
HRS & 100.00 & \textbf{100.00} & 100.00 & 91.18 & 100.00 \\
    & & & & & \\
    & \multicolumn{5}{c}{$p = 100$} \\
    & & & & & \\
BGS & 100.00 & 100.00 & 100.00 & 100.00 & 100.00 \\
HRS & 100.00 & 100.00 & 100.00 & 100.00 & 100.00 \\ \hline
    & & & & & \\
\multicolumn{6}{l}{\underline{MCC}} \\
    & & & & & \\
    & \multicolumn{5}{c}{$p = 30$} \\
    & & & & & \\
BGS & 7.85 & 12.39 & 5.19 & 3.45 & 11.74 \\
HRS & \textbf{48.17} & \textbf{52.89} & \textbf{36.44} & \textbf{50.90} & \textbf{61.08} \\ 
    & & & & & \\
    & \multicolumn{5}{c}{$p = 100$} \\
    & & & & & \\
BGS & 5.96 & 11.18 & 3.89 & 6.41 & 10.86 \\ 
HRS & \textbf{53.85} & \textbf{63.20} & \textbf{39.70}& \textbf{64.16} & \textbf{82.77} \\
\hline
Notes: & \multicolumn{5}{l}{(a) The better results are boldfaced.} \\
       & \multicolumn{5}{l}{(b) The figures are in percentages.}
\end{tabular}
\end{footnotesize}
\end{center}
\end{table}

To assess the performance of the graphical structure learning, we checked whether the point estimate of $\boldsymbol{\Omega}$ could successfully restore the true structure from the simulated data. Recall that there was no connection between nodes, e.g., node $i$ and node $j$ $(1\leqq i,\ j\leqq p)$, if $\omega_{ij} = 0$. Like Fan et al. (2009), we used the following rule to determine whether a pair of nodes was connected or not:
\begin{equation}
\label{threshold}
\begin{cases}
 |\hat\omega_{ij}| \geqq 10^{-3} & \text{(node $i$ and node $j$ are connected)}; \\
 |\hat\omega_{ij}| < 10^{-3} & \text{(node $i$ and node $j$ are not connected)},
\end{cases}
\end{equation}
where $\hat\omega_{ij}$ is the point estimate of $\omega_{ij}$ computed with the Monte Carlo sample of $\boldsymbol{\Omega}$ that we generated for each scenario with the HRS or BGS. Then, with the estimated graphical structures (50 in total), the accuracy in the graphical structure learning was measured with three criteria: specificity, sensitivity, and the Matthews correlation coefficient (MCC), namely
\begin{align}
\label{criteria}
\text{Specificity} &= \frac{\mathrm{TN}}{\mathrm{TN}+\mathrm{FP}}, \quad
\text{Sensitivity} = \frac{\mathrm{TP}}{\mathrm{TP}+\mathrm{FN}}, \nonumber \\
\text{MCC} &= \frac{\mathrm{TP}\times\mathrm{TN}-\mathrm{FP}\times\mathrm{FN}}
{\sqrt{(\mathrm{TP}+\mathrm{FP})(\mathrm{TP}+\mathrm{FN})(\mathrm{TN}+\mathrm{FN})(\mathrm{TN}+\mathrm{FN})}},
\end{align}
where TP, TN, FP, and FN are the number of true positives, true negatives, false positives, and false negatives, respectively, in the 50 replications.

Table 3 reports the calculated criteria\footnote{The results of the BGS in Table 3 are far different from those in Table 2. We assumed that this discrepancy was caused by the difference in the criteria for detecting connections. p882, Wang (2012) stated that ``we claim $\{\omega_{ij}=0\}$ if $\hat\omega_{ij} < 10^{-3}$ as Fan et al. (2009),'' which means that a negative $\hat\omega_{ij}$, whether near or far from 0, is regarded as evidence against a connection between nodes. As a result, negative relations between nodes would be over-rejected and the estimated graphical structure would be too sparse in the sense that the precision matrix would include too many zeros in the off-diagonal elements. To confirm this conjecture, we recalculated the three criteria in \eqref{criteria} without the absolute value in \eqref{threshold} and found that the recalculated results were comparably similar to those of Wang (2012).} for the 12 scenarios. As in Table 3, the HRS outperformed the BGS for all scenarios, except for the sensitivity of the Star model with $p=30$, though the sensitivity of the HRS was still over 90\%. Specifically, in the case of $p = 100$, the values of specificity were over 90\% for the HRS, which means that most of the zero off-diagonal elements in $\boldsymbol{\Omega}$ were correctly identified. This accuracy is crucial when trying to detect the true graphical structure in practice. It seems that imposing the positive definiteness constraint also enhanced the graphical structure learning in the Bayesian adaptive graphical LASSO.

\subsection{Application to S\&P500 Stock Return Data}
Next, we applied the BGS and HRS to stock return data and estimated $\Omega$. We used the standardized monthly excess return data against the S\&P 500 stock index for 483 stocks continuously listed from the end of December 2013 to the end of January 2018 of 505 constituents of the S\&P 500 as of February 2018 (n = 50, p = 483). The settings were the same as those for the simulation data. 

Although violation of the positive definiteness after updating the off-diagonal elements reached 808,009 times (16.73\%) in the BGS, it never occurred in the HRS. Figures 1 and 2 show the posterior mean of $\Omega$ by the BGS and HRS. Here, to make it easier to compare the BGS and HRS, we adjusted the scale of $\Omega$ so that the diagonal elements were one. The $\Omega$ estimated by the BGS in Figure 1 had many nonzero values remaining in the off-diagonal elements, while the off-diagonal elements of $\Omega$ estimated by the HRS in Figure 2 shrunk.

\begin{figure}[htbp]
\begin{tabular}{c}
\begin{minipage}[t]{0.45\hsize}
\centering
  \includegraphics[width=119mm]{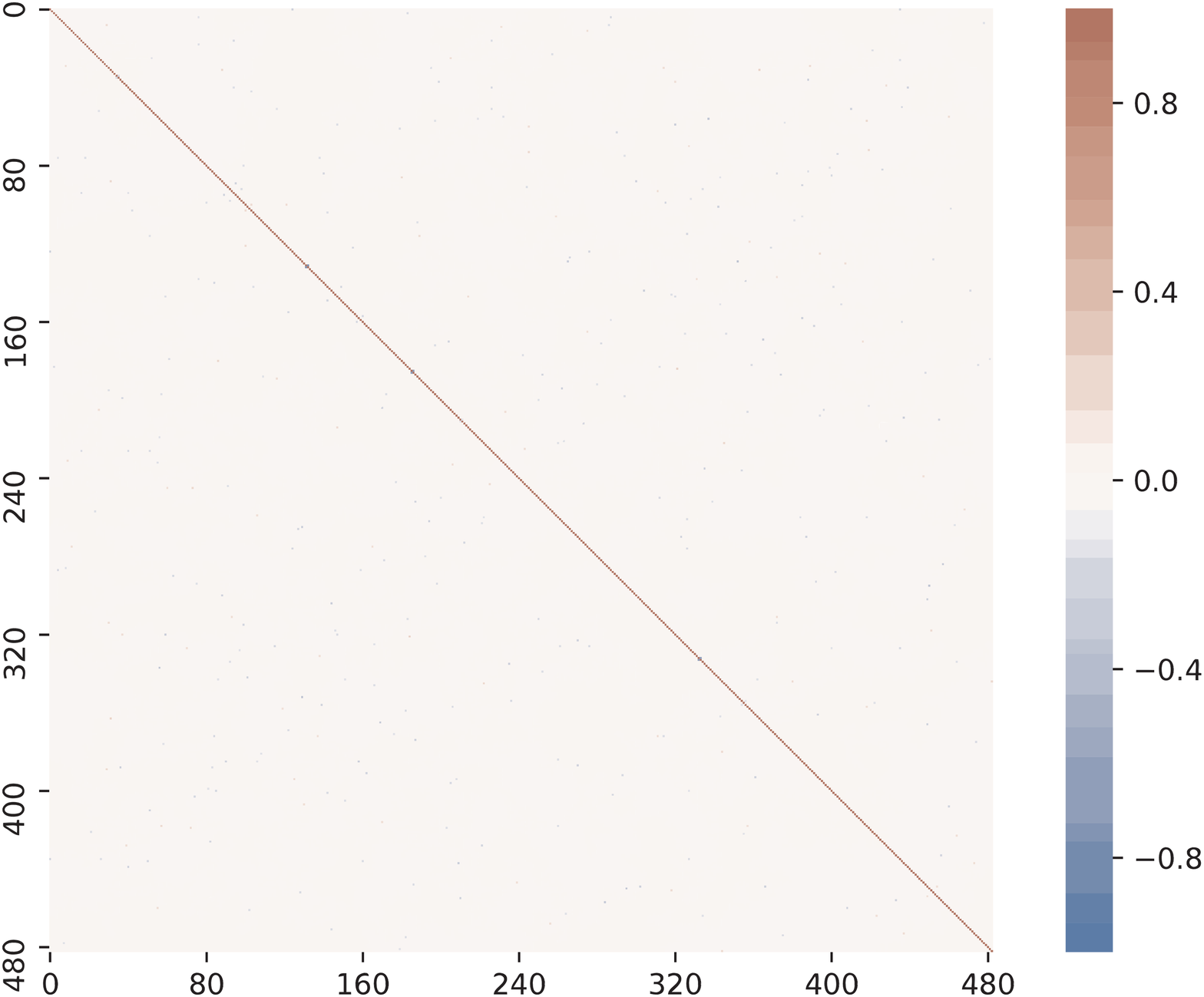}
\caption{Posterior mean of $\Omega$ by the BGS}
  \end{minipage} \\
  \begin{minipage}[t]{0.45\hsize}
\centering
  \includegraphics[width=119mm]{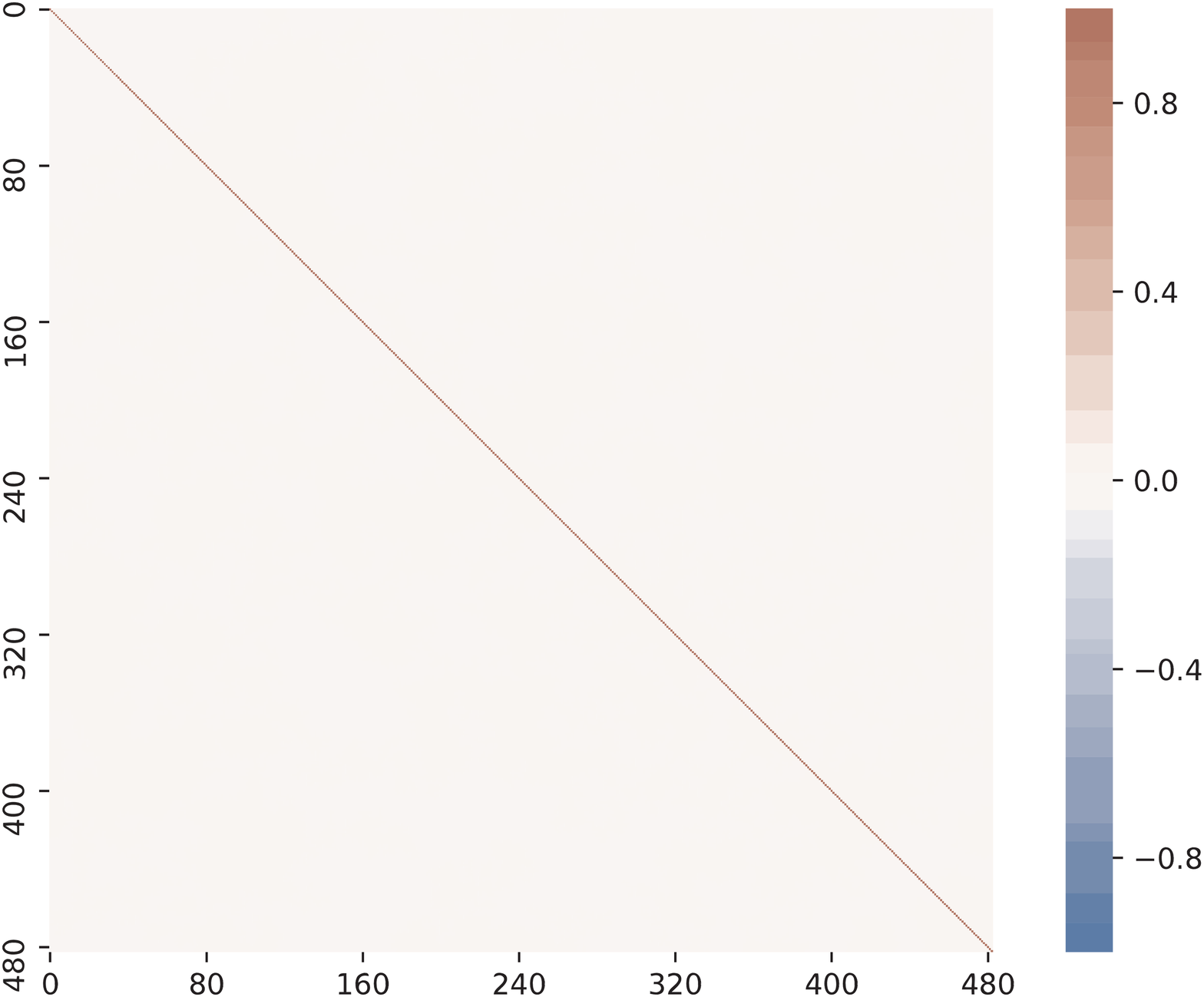}
\caption{Posterior mean of $\Omega$ by the HRS}
 \end{minipage}
\end{tabular}
\end{figure}

\section{Conclusion}
In this paper, we proposed a modification of Wang's (2012) block Gibbs sampling algorithm for the Bayesian graphical LASSO that we used as the primary example. Our modified algorithm guarantees the positive definiteness of the precision matrix throughout the sampling procedure by generating the off-diagonal elements of the precision matrix from a truncated multivariate normal distribution whose support is the region wherein the updated precision matrix remains positive definite. To facilitate sampling from such a complicated distribution, we proposed utilizing the hit-and-run algorithm by B\'{e}lisle et al. (1993). The derived algorithm is still a pure Gibbs sampler and maintains the efficiency and scalability of Wang's (2012) original algorithm. In the simulation study, we showed that our modified algorithm remarkably improved the accuracy in the point estimation and graphical structure learning. We also demonstrated that our modified algorithm could estimate the precision matrix even when the dimension of the precision matrix exceeds the sample size by applying it to the monthly return data of 483 stocks over 50 months. Since the key part of the Gibbs sampling algorithm in which the precision matrix is updated is common to other graphical models with shrinkage priors, such as the spike-and-slab prior (Wang [2015]), the horseshoe prior (Li et al. [2019]), and other scale-mixture-of-normals shrinkage priors, it would be simple to incorporate our modified algorithm into the Gibbs sampling algorithm for those models.

\section*{Supplementary Material}
\label{code_url}
Python codes for the block Gibbs sampler and the Hit-and-Run sampler used in this paper is available from \url{https://github.com/oyakeioecon/onglasso}.

\section*{acknowledgements}
This research is supported by JSPS Grants-in-Aid for Scientific Research Grant Number 19K01592 and the Keio Economic Society.

This preprint has not undergone peer review (when applicable) or any post-submission improvements or corrections. The Version of Record of this article is published in \textit{Japanese Journal of Statistics and Data Science}, and is available online at \burl{https://doi.org/10.1007/s42081-022-00147-1}.
%
\section*{Conflict of interest}
The authors declare that they have no conflict of interest.



\end{document}